\documentclass[12pt]{iopart}
\usepackage{amsfonts}
\usepackage{graphics}
\usepackage{graphicx}
\newcommand{\cN}{\mathcal{N}}
\newcommand{\dv}{\frac{\sfv_0'(z)-\sfv_0'(y)}{z-y}}
\newcommand{\sfv}{{\mathsf{v}}}
\newcommand{\tx}{{\tilde x}}
\newcommand{\tD}{{\tilde D}}
\newcommand{\ptx}{{\partial_{\tx}}}
\newcommand{\po}{{\mathsf{p}_1}}
\newcommand{\sfJ}{{\mathsf{J}}}
\newcommand{\elabel}[1]{\label{eq:#1}}
\newcommand{\gcomment}[1]{}

\begin{document}
\title{Orthogonal polynomials with discontinuous weights}

\author{Yang Chen$\dagger$ and Gunnar Pruessner$\ddagger$}

\address{$\dagger$Department of Mathematics,
Imperial College London,
180 Queen's Gate,
London SW7~2BZ,
UK}

\address{$\ddagger$Department of Physics,
Virginia Polytechnic Inst. \& State Univ.,
Blacksburg, VA 24061-0435,
USA}
\date{\today}

\ead{ychen@ic.ac.uk}

\begin{abstract}
In this paper we present a brief description of a ladder operator formalism
applied to orthogonal polynomials with discontinuous weights.
The two coefficient functions, $A_n(z)$ and $B_n(z)$,
appearing in the ladder operators satisfy the two
fundamental compatibility conditions previously derived for smooth weights.  If
the weight is a product of an absolutely continuous reference weight $w_0$
and a standard jump function, then $A_n(z)$ and $B_n(z)$ have apparent simple
poles at these jumps. We exemplify the approach by taking $w_0$ to be the Hermite
weight. For this simpler case we derive, without using the compatibility
conditions, a pair of difference equations satisfied by the diagonal and
off-diagonal recurrence coefficients for a fixed location of the jump. 
We also derive a pair of Toda evolution equations for the
recurrence coefficients which, when combined with the difference equations,
yields a particular Painlev\'e IV.
\end{abstract}

\submitto{\JPA}

\section{Introduction}
We begin this section by fixing the notation.  Let
$P_n(x)$ be monic polynomials of degree $n$ in $x$ and orthogonal, with respect
to a weight, $w(x),\;x\in[a,b]$, 
\begin{equation}
\int_{a}^{b} P_m(x) P_n(x) w(x) dx = h_n \delta_{m,n} \elabel{def_hn} \ ,
\end{equation}
where $h_n$ is the square of the weighted $L^2$ norm of $P_n.$ Also,
\begin{equation} \elabel{def_polys}
P_n(z) = z^n + \po(n) z^{n-1} + \dots \ .
\end{equation}
For convenience we set $w(a)=w(b)=0$. The recurrence relation follows from the 
orthogonality condition:
\begin{equation}
z P_n(z) = P_{n+1}(z) + \alpha_n P_n(z) + \beta_n P_{n-1}(z) \quad n=0,1,...
\elabel{recurrence_eq} \ ,
\end{equation}
where $\beta_0 P_{-1}(z):=0$, the
$\alpha_n,\;n=0,1,2,\dots$ are real and $\beta_n,\;n=1,2,\dots$ are strictly
positive. 

In this paper we describe a formalism which will facilitate the
determination of the recurrence coefficients for polynomials with singular
weights. Two points of view lead to this problem:
On one hand the X-ray problem \cite{BaCh} of condensed matter theory, 
on the other hand related problems in random matrix theory which involve the 
asymptotics of the Fredholm determinant of finite convolution operators with 
discontinuous symbols \cite{JimboETAL:1980}. 
This paper is the first in a series that
systematically study orthogonal polynomial where the otherwise smooth weights 
have been singularly deformed. The ultimate aim is the computation for large
$n$ of the determinant $D_n$ of the $n\times n$ moments or Hankel matrix
$(\mu_{j+k})_{j,k=0}^{n-1}$ with moments
$\mu_{j} := \int_{a}^{b} w(t) t^{j} dt$ where $j=0,1,\dots$,
thereby doing what has been done for the determinants 
of $n\times n$ Toeplitz matrices with singular generating functions \cite{BaWi}. 
 
The deformed weight with one jump is 
\begin{equation} \elabel{def_weight}
w(x) = w_0(x) \left( 1 - \frac{\beta}{2} + \beta \theta(x-\tx) \right) \ ,
\end{equation}
where $a<\tx<b$ is the position of the jump, $\theta(.)$ is the Heaviside step
function and the real $\beta$ parametrises the height of the jump.
More generally, we take $w_{\sfJ}>0$ to be the canonical jump function
$w_{\sfJ}(x) = 1 + \sum_{j=1}^{\cN} \Delta_j \theta(x-\tx_j)$
and $w(x) := w_0(x)w_{J}(x)$.

The actions of the ladder operators 
on $P_n(z)$ and $P_{n-1}(z)$ are
\begin{eqnarray}
\left( \frac{d}{dz} + B_n(z) \right) P_n(z) & = & \beta_n A_n(z) P_{n-1}(z) \elabel{ladder_Eq1} \\
\left( \frac{d}{dz} - B_n(z) - \sfv_0'(z) \right) P_{n-1}(z) & = &
             - A_{n-1}(z) P_n(z) \elabel{ladder_Eq2}
\end{eqnarray}
\begin{eqnarray} 
\fl
A_n(z) := \sum_{j=1}^{\cN} \frac{R_n(j)}{z-\tx_j} +
             \frac{1}{h_n} \int_{a}^{b} \dv P_n^2(y) w(y) dy
	     \elabel{ladder_An} && \\
\fl
B_n(z) := \sum_{j=1}^{\cN}\frac{r_n(j)}{z-\tx_j} + 
             \frac{1}{h_{n-1}} \int_{a}^{b}\dv P_{n-1}(y)P_n(y)w(y)dy 
	     \elabel{ladder_Bn} &&
\end{eqnarray}
where $w_0(x)=:\exp(-\sfv_0(x))$. If $w(a)$ and $w(b)$ are non-vanishing 
one must add
\begin{equation*}
\left. \frac{P_n^2(y) w(y)}{h_n(y-z)} \right|_{y=a}^{b}
\quad \textrm{and} \quad 
\left. \frac{P_n(y) P_{n-1}(y) w(y)}{h_{n-1}(y-z)} \right|_{y=a}^{y=b}
\end{equation*}
to \eref{eq:ladder_An} and \eref{eq:ladder_Bn} respectively.

Now $A_n(z)$ and $B_n(z)$, the coefficient functions in the ladder operators,
satisfy identities analogous to those found for smooth weights 
\cite{ChIs1,IW,ChIs3}:
\begin{eqnarray}
B_{n+1}(z) + B_n(z) & = & (z-\alpha_n) A_n(z) -\sfv_0'(z) \elabel{S_1} \\
B_{n+1}(z) - B_{n}(z) & = & \frac{\beta_{n+1} A_{n+1}(z) - \beta_n A_{n-1}(z) - 1}{z-\alpha_n}
\elabel{S_2} \\
R_n(\tx_j) = R_n(\Delta_1,...,\Delta_{\cN},\tx_j) & := & 
   \frac{\Delta_j}{h_n} P_n^2(\tx_j) w_0(\tx_j)
   \elabel{def_Rn_general}\\
r_n(\tx_j) = r_n(\Delta_1,...,\Delta_{\cN},\tx_j) & := &
   \frac{\Delta_j}{h_{n-1}} P_n(\tx_j) P_{n-1}(\tx_j) w_0(\tx_j) 
   \elabel{def_rn_general} 
\end{eqnarray}
The derivation of \eref{eq:ladder_Eq1}-\eref{eq:def_rn_general} 
will be published in a forthcoming paper where the weight has several
jumps and $w_0(x)=(1-x)^{\alpha}(1+x)^{\beta},\;\;x\in[-1,1]$ is the Jacobi
weight.  

Multiplying the recurrence relation \eref{eq:recurrence_eq} evaluated at $z=\tx_j$ by 
$\Delta_jw_0(\tx_j)P_n(\tx_j)$ and noting \eref{eq:def_Rn_general} as well as
\eref{eq:def_rn_general} we arrive at the universal equality
\begin{equation}
(\tx_j-\alpha_n) R_n(\tx_j) = r_{n+1}(\tx_j)+r_n(\tx_j) \ .\elabel{U_1}
\end{equation}
Similarly, squaring $r_n(\tx_j)$ we find a second universal equation
\begin{equation}
r_n^2(\tx_j)=\beta_nR_n(\tx_j)R_{n-1}(\tx_j) \ .\elabel{U_2}
\end{equation}

Note that in the expressions for $A_n(z)$ and $B_n(z)$ only $\sfv_0$, the
``potential'' associated with the smooth reference weight, appears. The
discontinuities give rise to $R_n(\tx_j)$ and $r_n(\tx_j).$

It is clear from \eref{eq:ladder_An} and \eref{eq:ladder_Bn} that
if $\sfv_0'(z)$ is rational, then $A_n(z)$ and $B_n(z)$
are also rational. This is particularly useful for our purpose which is the 
determination of the recurrence coefficients, for in this situation by
comparing residues on both sides of \eref{eq:S_1} and \eref{eq:S_2} we should find
the required difference equations \cite{ChIs3}.

In the following section the above approach is exemplified by the Hermite
weight, $w_0(x)=\exp(-x^2)$ and $w$ given by \eref{eq:def_weight}. It turns out
that in this situation $\alpha_n$ and $\beta_n$ are related to $R_n$ and
$r_n$ in a very simple way.

\section{Hermite weight with one jump}
Now, $w_0(x)=\exp(-x^2)$, so that $\sfv_0(x)=x^2$, and $w(x)$ as in
\eref{eq:def_weight}. Also,
\begin{eqnarray}
R_n & := & R_n(\tx)=\beta \frac{P_n^2(\tx,\tx)}{h_n(\tx)} w_0(\tx)
\elabel{def_Rn} \\
r_n & := & r_n(\tx)=\beta \frac{P_n(\tx,\tx) P_{n-1}(\tx,\tx)}{h_{n-1}(\tx)} 
w_0(\tx) \elabel{def_rn} \ ,
\end{eqnarray}
which are independent of the particular choice of $w_0$ and 
\begin{equation}
A_n(z) =  \frac{R_n}{z-\tx} + 2 \elabel{ABn_Hermite} \quad \textrm{and} \quad
B_n(z) =  \frac{r_n}{z-\tx} \ ,
\end{equation}
particular to $\sfv_0(x)=x^2.$ Note that $P_j(\tx,\tx)$ is the value of
$P_j(z,\tx)$ at $z=\tx.$

Instead of proceeding with the full machinery
of \eref{eq:S_1} and \eref{eq:S_2} we take advantage of the fact that $\sfv_0(x)=x^2$.
From orthogonality and the recurrence relation, we have
\begin{equation} \elabel{an_hn}
\alpha_n(\tx) h_n(\tx) = \frac{1}{2}
\int_{-\infty}^{\infty}\sfv_0'(x) P^2_n(x,\tx) w(x) dx =
\frac{R_n(\tx)}{2} h_n(\tx) 
\end{equation}
by integration by parts. The string equation,
\begin{equation}
\elabel{string_equation}
n h_{n-1}(\tx) = 
  \int_{-\infty}^{\infty} P_n'(x,\tx) P_{n-1}(x,\tx) w(x) dx \ ,
\end{equation}
is an immediate consequence of the orthogonality condition.
Again, an integration by parts and noting that $\beta_n = h_n/h_{n-1}$
produces
\begin{equation} \elabel{beta_n_from_r_r_n}
\beta_n(\tx) = \frac{n}{2} + \frac{r_n(\tx)}{2} \ .
\end{equation}
It should be pointed out here that in general neither the string equation
\eref{eq:string_equation} nor \eref{eq:an_hn} will provide the complete set of
difference equations for the recurrence coefficients which can be seen if
$w_0$ were the Jacobi weight. In such a situation the compatibility conditions
\eref{eq:S_1} and \eref{eq:S_2} must be used. 

Now \eref{eq:U_1} and \eref{eq:U_2} become
\begin{equation} \elabel{diff1}
\frac{r_{n+1} + r_n}{2}=\alpha_n (\tx - \alpha_n) \ .
\end{equation}
and
\begin{equation} \elabel{diff2}
r_n^2 = 2 (n + r_n) \alpha_{n} \alpha_{n-1} \ .
\end{equation}
Equations \eref{eq:diff1} and \eref{eq:diff2}, supplemented by the initial conditions
\begin{equation*}
\fl
\alpha_0(\tx) = \frac{\beta}{2} \exp(-\tx^2)
\left[
(1-\beta/2) \sqrt{\pi} + \beta \int_{\tx}^{\infty} \exp(-t^2)dt 
\right]^{-1}
\quad \textrm{and} \quad
r_0(\tx) = 0 \ ,
\end{equation*}
can be iterated to determine the recurrence coefficients numerically. 
Also, 
explicit solutions to \eref{eq:diff1} and \eref{eq:diff2} can be produced for
small $n$.

\section{Derivative with respect to $\tx$ and Painlev\'e IV}
If \eref{eq:diff1} and \eref{eq:diff2} are combined with the evolution
equations to be derived in this section, the Painlev\'e IV mentioned in
the abstract is found.
We begin with the $L^2$ norm $h_n(\tx)$, \Eref{eq:def_hn}, 
which entails 
\begin{equation}
\fl
\partial_{\tx} h_n =  -\beta \int_{-\infty}^{\infty} P_n^2(x) w_0(x)
                          \delta(x-\tx) dx 
                   =  -\beta P_n^2(\tx) w_0(\tx) = - h_n R_n = -2 h_n \alpha_n 
		   \elabel{h_Rn_alpha} \ .
\end{equation}
and thus
$\partial_{\tx} \ln \beta_n = -2 (\alpha_n - \alpha_{n-1})$
since $\ln \beta_n=\ln h_n-\ln h_{n-1}$. With \eref{eq:beta_n_from_r_r_n}, 
\begin{equation} \elabel{first_Toda}
\frac{1}{2 (n+r_n)} \ptx r_n = \alpha_{n-1}-\alpha_n
\end{equation}
which is the first Toda equation. 
Taking the derivative with respect to $\tx$ of \eref{eq:def_hn} at $m=n-1$ and 
using the definition of the
monic polynomials \eref{eq:def_polys} then gives
$\partial_{\tx} \po(n,\tx) = r_n$
since $\po(n,\tx) - \po(n+1,\tx)=\alpha_n(\tx)$ is an immediate 
consequence of the recurrence relation. Therefore
\begin{equation} \elabel{second_Toda}
\partial_{\tx} \alpha_n=r_n-r_{n+1} \ ,
\end{equation}
the second Toda equation.

Eliminating $r_{n+1}$ from \eref{eq:diff1} and the second Toda equation
\eref{eq:second_Toda}, gives $r_n$ in terms of $\alpha_n$ and $\ptx \alpha_n$:
\begin{equation} \elabel{DE_alpha_to_rn}
r_n = \alpha_n(\tx - \alpha_n) + \frac{1}{2} \ptx \alpha_n \ .
\end{equation}
Using the first Toda equation \eref{eq:first_Toda} to express $\alpha_{n-1}$ in
terms of $\alpha_n$ and $\ptx r_n$ and substituting \eref{eq:DE_alpha_to_rn}
into \eref{eq:diff2} produces a particular Painlev\'e IV \cite{Bassom}, 
\begin{equation} \elabel{painleve}
\alpha_n'' = \frac{(\alpha_n')^2}{2 \alpha_n} + 6 \alpha_n^3 
             - 8 \tx \alpha_n^2 + 2(\tx^2 - (2n+1)) \alpha_n \ ,
\end{equation}
which can be brought into the canonical form 
with the replacements $\alpha_n\to \alpha_n/2$ and $\tx\to -\tx$.
\Eref{eq:painleve} is supplemented by the boundary conditions
$\lim_{\tx \to \pm \infty} \alpha_n(\tx) = 0$. In a recent paper
\cite{Forr}, a Painlev\'e IV was derived for the discontinuous Hermite weight
using an entirely different method.

Based on \eref{eq:h_Rn_alpha} and \eref{eq:second_Toda} the derivative of the 
logarithm of the Hankel determinant $D_n=\prod_{j=0}^{n-1}h_j$ can be computed
as
\begin{eqnarray}
\partial_{\tx} \ln D_n    & = &
       -2 \sum_{j=0}^{n-1} \alpha_j = -\sum_{j=0}^{n-1}R_j 
       = -\beta \exp(-\tx^2) \sum_{j=0}^{n-1} \frac{P_j^2(\tx,\tx)}{h_j(\tx)}
       \\
\partial_{\tx}^2 \ln D_n  & = & 2 r_n
\elabel{d2_ln_Hankel} \ .
\end{eqnarray}
where \eref{eq:def_Rn} has been used in the first line, which can be summed 
by the Christoffel-Darboux formula,
\begin{equation*} 
\sum_{j=0}^{n-1} \frac{P_j(x,\tx)P_j(y,\tx)}{h_j(\tx)}
=\frac{P_n(x,\tx)P_{n-1}(y,\tx)-P_n(y,\tx)P_{n-1}(x,\tx)}{(x-y) h_{n-1}(\tx)}
\ .
\end{equation*}
In the limit $x\to y,$ we find, in general, 
\begin{eqnarray*} 
h_{n-1}(\tx) \sum_{j=0}^{n-1} \frac{P_j^2(x,\tx)}{h_j(\tx)} && \\
\fl
= \beta_n A_n(x) P_{n-1}^2(x,\tx) + A_{n-1}(x) P_n^2(x,\tx)
   -\big( 2 B_n(x) + \sfv_0'(x) \big) P_n(x,\tx) P_{n-1}(x,\tx) &&
\end{eqnarray*}
using the the ladder operators \eref{eq:ladder_Eq1} and \eref{eq:ladder_Eq2}.
With \eref{eq:ABn_Hermite} this entails 
\begin{eqnarray*} \fl
\sum_{j=0}^{n-1}\frac{P_j^2(x,\tx)}{h_j(\tx)}
    =  \frac{1}{h_{n-1}(\tx)} 
              \left( 2 \beta_n P_{n-1}^2(x,\tx) + 2 P_n^2(x,\tx)
                - 2 x P_n(x,\tx) P_{n-1}(x,\tx) \right) && \\
\fl \qquad   +   \frac{1}{(x-\tx) h_{n-1}(\tx)}
              \left( \beta_n R_n P_{n-1}^2(x,\tx) + R_{n-1} P_n^2(x,\tx) 
	        - 2 r_n P_n(x,\tx) P_{n-1}(x,\tx)\right) &&
\end{eqnarray*}
The apparent pole at $x=\tx$ can be shown to have vanishing residue by
considering $\beta \exp(-\tx^2) \times (\textrm{residue})$:
\begin{eqnarray*} \fl
\beta \exp(-\tx^2) \Big( \beta_n R_n P_{n-1}^2(\tx,\tx) + R_{n-1} P_n^2(\tx,\tx) -
     2 r_n P_n(\tx,\tx) P_{n-1}(\tx,\tx) \Big)/h_{n-1} && \\
\fl
= 2 (n + r_n) \alpha_{n} \alpha_{n-1} + 2 (n+r_n) \alpha_{n-1}\alpha_n - 2 r_n^2
= 2 \big( 2 (n+r_n) \alpha_n \alpha_{n-1} - r_n^2\big) = 0 && \ ,
\end{eqnarray*}
where the last equality is due to \eref{eq:diff2}.
A further regular term can be found as a contribution from 
the Taylor series of $P_j(x,\tx)$ about $x=\tx$  
namely, 
\begin{eqnarray}
&& 2 \frac{\beta_n}{h_{n-1}} R_n P_{n-1}(\tx,\tx) P_{n-1}'(\tx)
 + 2 \frac{R_{n-1}}{h_{n-1}} P_n(\tx,\tx) P_n'(\tx) \nonumber \\
&& - 2 \frac{r_n}{h_{n-1}} \partial_x 
               \left. \left(P_{n-1}(x,\tx) P_n(x,\tx) \right)
 \right|_{x=\tx} 
\elabel{second_reg_term}
\end{eqnarray}
where $P_j'(\tx) := \ptx P_j(x,\tx)|_{x=\tx}$. 
Using the fact that $\beta_n = h_n/h_{n-1}$ and \eref{eq:def_Rn},
we see that the first two terms of \eref{eq:second_reg_term} combined 
into $2(r_n/h_{n-1}) \partial_x (P_n(x,\tx)P_{n-1}(x,\tx))|_{x=\tx}$ cancel the 
third. We are therefore left with the regular term: 
\begin{eqnarray} \fl
\ptx \ln D_n =  -\beta  \left(2 \beta_n \frac{P_{n-1}^2(\tx,\tx)}{h_{n-1}}
                                 + 2 \frac{P_n^2(\tx,\tx)}{h_{n-1}}
				 - 2 \tx \frac{P_n(\tx,\tx)P_{n-1}(\tx,\tx)}{h_{n-1}}
			    \right) \exp(-\tx^2) \nonumber && \\
              =  2 \tx r_n - 2 (n+r_n)(\alpha_n+\alpha_{n-1}) 
	           \elabel{simple_d1_ln_Hankel} \ . &&
\end{eqnarray}
Using \eref{eq:first_Toda}--\eref{eq:DE_alpha_to_rn} 
it follows from 
$
2 \tx r_n' - 2r_n' (\alpha_n + \alpha_{n-1}) 
           - 2 (n + r_n) (\alpha_n' + \alpha_{n-1}')
 =  4 (n + r_n) \left( r_{n-1} - r_n - \alpha_{n-1}' \right) = 0
$
that \eref{eq:simple_d1_ln_Hankel} reproduces the second derivative 
\eref{eq:d2_ln_Hankel}, $\partial_{\tx}^2 \ln D_n  = 2 r_n$.

Let $F_n(\tx) := -\ln D_n(\tx)$ be the free energy. Expressing
\eref{eq:simple_d1_ln_Hankel} in terms of
$\alpha_n(\tx)$ and finally in terms of $\Psi_n(\tx)$, where $\alpha_n(\tx) =:
\frac{\beta}{2}(\Psi_n(\tx))^2$, the free energy reads
\begin{eqnarray*} \fl
F_n(\tx)-F_n(-\infty)
  & = & \frac{\beta}{2} \int_{-\infty}^{\tx}
    \left( (4 n + 1 - 2 x^2) \Psi_n^2(x) + 3 \beta x \Psi_n^4(x) 
           - \beta^2 \Psi^6_n(x) \right) dx \\
  &   & +\frac{\beta}{2} \Psi_{n}(\tx) \Psi_n'(\tx) \ ,
\end{eqnarray*}
where $F_n(-\infty)$ is the free energy corresponding to $w(x)=\exp(-x^2)$.
Note that $\exp(-F_n(\pm\infty)) = 2\pi^{n/2} \prod_{k=1}^{n}
\frac{\Gamma(k)}{2^k}$, which gives rise to the sum rule
\begin{equation*}
\int_{-\infty}^{\infty} \Big( (4n + 1 - 2 x^2) \Psi_n^2(x) 
                           + 3 \beta x \Psi_n^4(x)
                           - \beta^2  \Psi^6_n(x) \Big) dx = 0 \ .
\end{equation*}

With a minor change of variables \eref{eq:d2_ln_Hankel} becomes the Toda 
molecule equation. First we 
note that
\begin{equation*}
r_n(\tx)      =  2 \beta_n(\tx) - n  \quad \textrm{and} \quad
\beta_n(\tx)  =  \frac{h_n(\tx)}{h_{n-1}(\tx)} 
         = \frac{D_{n+1}(\tx) D_{n-1}(\tx)}{(D_n(\tx))^2} 
\ .
\end{equation*}
Defining $D_n(\tx) =: \exp(-n\tx^2)\tD_n(\tx)$ it then follows 
\begin{equation*}
\ptx^2 \ln \tD_n = 4 \frac{\tD_{n+1} \tD_{n-1}}{(\tD_n)^2} \ .
\end{equation*}
We may express $\alpha_n(\tx)$ in terms of the derivatives of the
free energy, by noting that
$-F_n'         =  2 \tx r_n - 2 (n+r_n) (\alpha_n + \alpha_{n-1})$
and
$-F_n''        =  2 r_n$.
One finds with \eref{eq:first_Toda}
\begin{equation*}
\alpha_n(\tx)  = 
     \frac{F_n' - \tx F_n'' + F_n'''/2}{4 n - 2 F_n''} 
     = \frac{\tx}{2} - \frac{1}{4} \frac{f_n''' + 2 f_n'}{f_n''}
\end{equation*}
where $F_n(\tx) =: f_n(\tx) + n \tx^2$.

\section{Asymptotics and Numerics}
\begin{figure}[tb]
\begin{center}
\scalebox{0.5}{
\includegraphics[clip=true]{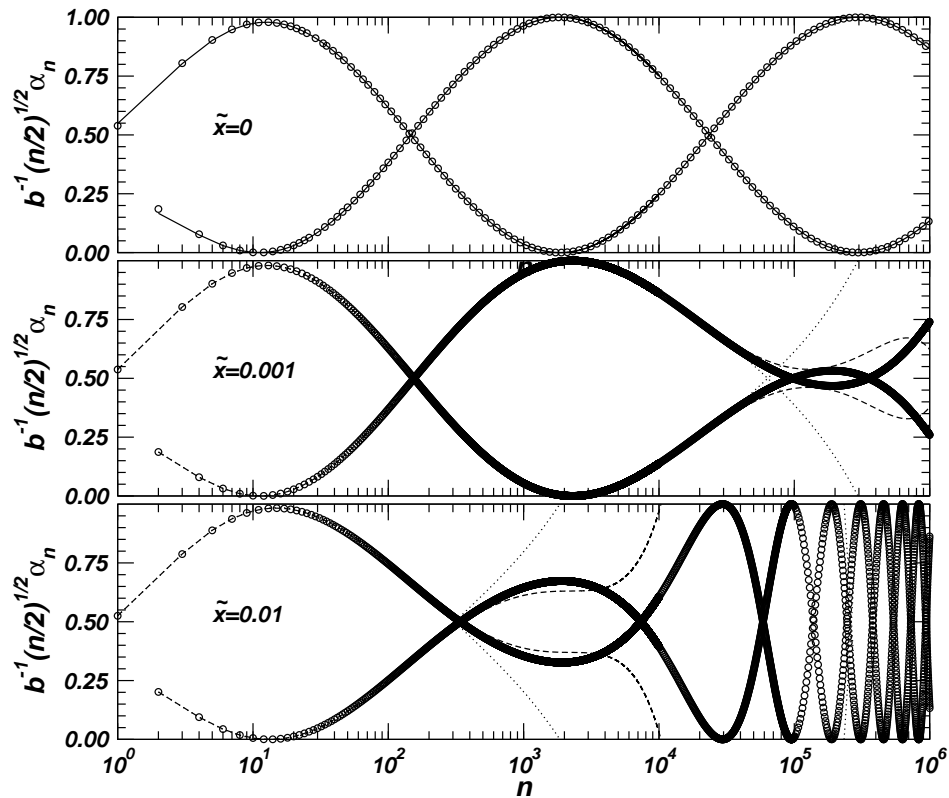}}
\caption{\label{fig:asymptote_nonzero_x}
Behaviour of $b^{-1} (n/2)^{1/2} \alpha_n$ for $\beta=3/2$ and various (small) values of $\tx$ 
from a numerical iteration of \eref{eq:diff1} and \eref{eq:diff2}. The top 
panel shows additionally \Eref{eq:asymptote_alpha} as a solid line passing
through the open circles. The
other two panels show the approximation of the rescaled $\alpha_n(\tx)$ using
the first three (two) terms of a Taylor series around $\tx=0$ as dashed
(dotted) line. The
behaviour of $r_n$ looks likewise. The numerical data is pruned differently in
the three cases to avoid artefacts.}
\end{center}
\end{figure}

For $\tx=0$ we find the asymptotic expansion
\begin{eqnarray}
\fl \alpha_n(0) & = & \frac{b}{\sqrt{2 n}} 
  \Big[ 1 + (-1)^n \sin(2 b \ln n + B)    \elabel{asymptote_alpha} \\
\fl   &  &  -   \frac{1}{4 n}\big(1 + (-1)^n (\sin(2 b \ln n + B) - 
	                            4 b \cos(2 b \ln n + B)) \big)
       + \mathcal{O}(n^{-2}) \Big] \nonumber \\ 
\fl r_n(0) & = & -b (-1)^n \cos( 2 b \ln n + B) - 
              \frac{b^2}{2n} \Big( 1+\sin^2(2b\ln n+B) \Big) 
	      + \mathcal{O}(n^{-2})  \elabel{asymptote_rn}
\end{eqnarray}
guided by the numerics on the difference equations \eref{eq:diff1} and
\eref{eq:diff2}. The constant $b$ is given by 
\begin{equation*}
b := \frac{1}{2 \pi} \ln \left(\frac{1 + \beta/2}{1 - \beta/2}\right)
\quad \textrm{with} \quad 
\frac{\beta}{2}\in(-1,1)
\end{equation*}
and $B$ is a phase independent of $n$. Unfortunately, the formalism developed
in this paper does not seem to shed any light on its delicate dependence on $\beta$.

\begin{figure}[tb]
\begin{center}
\scalebox{0.5}{
\includegraphics[clip=true]{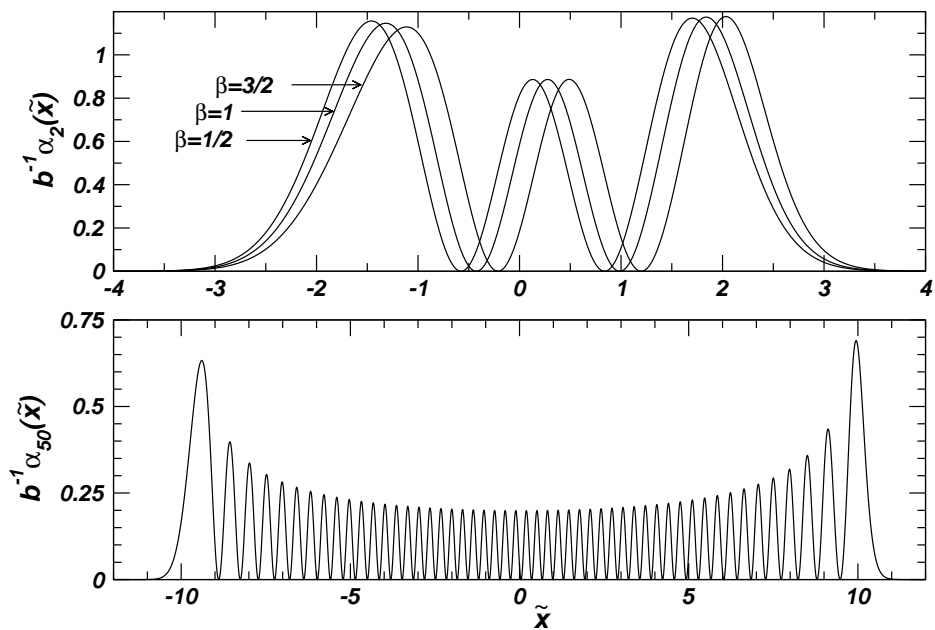}}
\caption{\label{fig:alpha_in_x_twiddle}
Upper panel: $b^{-1} \alpha_2(\tx)$ for different jump heights $\beta$ as a
function of $\tx$. 
Lower panel: For large $n$ the rescaled coefficient $b^{-1} \alpha_{50}(\tx)$
fluctuates wildly.}
\end{center}
\end{figure}

It can be verified by a direct calculation
that \eref{eq:asymptote_alpha} and \eref{eq:asymptote_rn} satisfy
\eref{eq:diff1} and \eref{eq:diff2} to order $1/n^2$. The top panel of
\fref{fig:asymptote_nonzero_x}
shows a comparison between the numerical results and the above asymptotes for
suitably rescaled $\alpha_n$. In principle, $\alpha_n$ and $r_n$ can be
determined analytically to any order in $n$ because an approximation of
$\alpha_n$ to order $n^{-m}$ gives rise to a difference equation for $r_n$ to
order $n^{-(m+1)}$ via \eref{eq:diff1}. In turn, $r_n$ to order $n^{-(m+1)}$
produces an equation for $\alpha_n$ to order $n^{-(m+1)}$ and so forth. This
scheme breaks down for $\tx\neq 0$.

For fixed $\tx \neq 0$, the numerics does not suggest an ansatz for the
asymptotes. Most remarkably, the effect of $\tx \neq 0$ persists for very
large $n$, even for very small $\tx \neq 0$, as illustrated in
\fref{fig:asymptote_nonzero_x}. Also shown in this figure as dashed (dotted)
lines are the approximations of $\alpha_n(\tx)$ from the first three (two)
terms of a Taylor-series in $\tx$ around $\tx=0$ based on the iterative
results for $\alpha_n(\tx=0)$, \eref{eq:second_Toda} and \eref{eq:painleve}.
In principle, the Painlev\'e IV, \Eref{eq:painleve}, provides a way to express
$\alpha_n^{(m)}(\tx)$ in terms of lower order derivatives $\alpha_n^{(m')}(\tx)$
with $m'<m$, yet the results in \fref{fig:asymptote_nonzero_x} suggest that
for sufficiently large $n$ a finite Taylor series eventually deviates wildly
from the correct $\alpha_n(\tx)$. Note that both $\sqrt{n} \alpha_n(\tx)$ and
$r_n(\tx)$ are bounded in $n$ for large $n$. 

\Fref{fig:alpha_in_x_twiddle} shows the rescaled $\alpha_n$ for fixed $n$ and
varying $\tx$. It resembles a Hermite polynomial because of its direct
relation to $P_n^2(\tx,\tx) w_0(\tx)$, \eref{eq:an_hn} and \eref{eq:def_Rn}.
For the same reason $\alpha_n(\tx)$ vanishes for $n \ln(\tx) \ll \tx^2/2$.

\ackn{
GP would like to thank EPSRC, the NSF (DMR-0088451/0414122) and the Humboldt Foundation for their
generous support.
}


\section*{References}
\begin{thebibliography}{xxx}
\bibitem{BaCh} E. L. Basor and Y. Chen, ``The X-ray problem revisited'',
J. Phys. A.: Math. Gen. {\bf 36} (2003) L175-L180; ``A note on the Wiener-Hopf
determinants and the Borodin-Okunkov identity'', Integr. Equ. Oper. Theory,
{\bf 45} (2003) 301-308.
\bibitem{BaWi} H. Widom, ``Toeplitz determinants with singular generating
functions'',
Amer. J. Math., {\bf 95} (1973) 333-383; E. L. Basor, ``Asymptotic formulas
for Toeplitz determinants'', Trans. Amer. Math. Soc. {\bf 239} (1978) 33-65.
\bibitem{Bassom} A. P. Bassom, P. A. Clarkson, A. C. Hicks, and J. B.
McLeod, ``Integral equations and exact solutions of the fourth Painlev\'e 
equation'',
Proc. R. Soc. London Ser. A {\bf 437} (1992) 1-24.
\bibitem{ChIs3} Y. Chen and M. E. H. Ismail, ``Jacobi polynomials from compatibility
conditions'', Proc. Amer. Math. Soc. {\bf 133} (2005) 465-472,
225-237.
\bibitem{ChIs1}Y. Chen and M. E. H. Ismail, ``Ladder operators
and differential equations for orthogonal polynomials'',
J. Phys. A.: Math. Gen. {\bf 30} (1997) 7817-7829.
\bibitem{Forr} P. J. Forrester and N. S. Witte, ``Discrete Painlev\'e
equations and random matrix averages'', Nonlinearity, {\bf 16} (2003)
1919-1944.
\bibitem{IW}M. E. H. Ismail and J. Wimp, ``On differential
equations for orthogonal polynomials'', Methods Appl. Anal.
{\bf 5} (1998) 439-452.
\bibitem{JimboETAL:1980}M. Jimbo, T. Miwa, Y. M\^ori and M. Sato, ``Density
matrix of an impenetrable Bose gas and the fifth Painlev\'e transcendent'', Physica D {\bf
1} (1980) 80--158.
\end{thebibliography}
\end{document}